\documentclass[%
 reprint,
 superscriptaddress,
 amsmath,amssymb,
 aps,
prb,
floatfix,
longbibliography 
]{revtex4-2}

\usepackage[pdftex]{graphicx} \graphicspath{{}}% Include figure files
\usepackage{float}
\usepackage{dcolumn}% Align table columns on decimal point
\usepackage{paralist}
\usepackage{enumitem} %nested lists
\usepackage{dsfont}
\usepackage{ulem} % strike out
\usepackage{bm}% bold math
\usepackage{mathtools} %mathclap
\usepackage{xcolor} %for color gray
\usepackage[pdftex,colorlinks=true]{hyperref}
\hypersetup{
  colorlinks=true,
  linkcolor=blue,
  urlcolor=cyan,
}

\usepackage{color}
\definecolor{ForestGreen}{RGB}{34, 139, 34}

\newcommand{\prbrev}[1]{{ #1}}

\newcommand{\unue}[1]{\textit{#1}}

\def\closedsqrt{\mathpalette\DHLhksqrt}
\def\DHLhksqrt#1#2{%
\setbox0=\hbox{$#1\sqrt{#2\,}$}\dimen0=\ht0
\advance\dimen0-0.2\ht0
\setbox2=\hbox{\vrule height\ht0 depth -\dimen0}%
{\box0\lower0.4pt\box2}}

\begin{document}

\preprint{APS/123-QED}

\title{Dynamical correlations  and domain wall relocalization in transverse field Ising chains}% Force line breaks with \\

\author{Philippe Suchsland}
\affiliation{Max Planck Institute for the Physics of Complex Systems, N\"{o}thnitzer Str. 38, 01187 Dresden, Germany}

\author{Beno\^{\i}t Dou\c{c}ot}
 \affiliation{Max Planck Institute for the Physics of Complex Systems, N\"{o}thnitzer Str. 38, 01187 Dresden, Germany}
 \affiliation{LPTHE, UMR 7589, CNRS and Sorbonne Universit\'e, 75252 Paris Cedex 05, France}
 
\author{Vedika Khemani}
 \affiliation{Department of Physics, Stanford University, Stanford, California 94305, USA}

\author{Roderich Moessner}
 \affiliation{Max Planck Institute for the Physics of Complex Systems, N\"{o}thnitzer Str. 38, 01187 Dresden, Germany}

\begin{abstract}
\noindent
We study conventional and out-of-time-ordered correlators (OTOCs) for a wide variety of transverse field Ising chains: classical and quantum, clean and disordered, and integrable and generic. The setting we consider is that of a quantum quench. We find a remarkably rich phenomenology,  ranging from stable periodic signals to ones decaying with varying rates. This variety is due to a complex interplay of  constraints  on thermalization imposed by integrability and symmetry. A process we term dynamical domain wall relocalization provides a long-lived signal in the clean, integrable case, which can be degraded by the addition of disorder even without interactions. Our results shed light on a proposal to use an OTOC as a  dynamical diagnostic of a quantum phase  more powerful than a standard observable.
\end{abstract}

\maketitle

\unue{Introduction.} %
The advent of experiments on coherently evolving quantum matter has led to an interdisciplinary focus on quantum dynamics~\cite{2_1_bloch2008,2_2_bloch2012,2_3_blatt2012,2_4_georgescu2014,2_5_schreiber2015,2_6_smith2016,2_7_bordia2016,2_8_choi2016,2_9_martinez2016,2_10_flaeschner2017,2_11_jurcevic2017,2_12_zhang2017,2_13_zhang2017,2_14_choi2017,2_X_1_joshi2020,60_swan2019,Lenar_i__2018}. In contrast with conventional time-ordered dynamical correlation functions of two operators $O_{p,m}$, {$\langle O_\mathrm{p}(t) O_\mathrm{m}\rangle$}, several recent studies have focused on out-of-time-ordered correlators (OTOCs), {$\langle O_\mathrm{p}(t) O_\mathrm{m} O_\mathrm{p}(t) O_\mathrm{m}\rangle$}. Proposed by Larkin \textit{et al}. \cite{larkin_otoc,2_23_li2017,2_24_gaerttner2017,2_25_yao2016,2_26_swingle,2_27_zhu2016,2_28_campisi2017,2_29_aleiner2016,2_30_bohrdt2017,2_31_fan2017,2_32_huang2016}, OTOCs probe how a system retains or loses memory of its initial state when subject to a perturbation, and have been studied in relation to a variety of phenomena ranging from quantum chaos to black hole physics to operator spreading and information scrambling~\cite{2_33_maldacena2016,60_2_hayden2007,2_X_2_sekino2008,0_von_Keyserlingk_2018,1_khemani2018,Rakovszky_2018,otoc_floquet_dyn_phase_trans}. 

One recent line of inquiry concerns the possible capacity of OTOCs to act as diagnostics for zero-temperature phases in quench experiments in a way which is not accessible to traditional time-ordered correlators~\cite{heylpollmann,2_X_sun_otoc,2_X_2_wang2019,2_X_1_sun2020,2_X_16_shen2017,2_X_X_nie2020,2_X_X_chen_2020}. A case in point is the  one-dimensional transverse field Ising model (TFIM), where recent work provided numerical evidence that the nonvanishing  OTOC of an order parameter (concretely, the local magnetization) evaluated at late times detects the presence of ground-state ordering, even starting from an initially fully polarized state~\cite{heylpollmann,2_X_sun_otoc,2_X_X_nie2020,2_X_X_chen_2020}. The numerics indicated this behavior for both the TFIM, which is integrable, and a perturbed (interacting/generic) nonintegrable model. 
These results were surprising because the polarized state is a finite temperature state for any nonzero transverse field, and clean one-dimensional systems do not order at finite temperature.
Indeed, in contrast with the OTOC, standard temporal correlators of the magnetization  decay to zero, as expected from thermalization to a paramagnetic finite temperature equilibrium state. 
This raises the theoretical question of how the OTOC might manage to evade thermalization to detect the order present in the ground-state of the Hamiltonian, but absent in equilibrium at the energy density corresponding  to  the initial state of the quench. 

In our work, we resolve this and other questions,  providing a detailed and comprehensive study of OTOCs for a variety of Ising chains. We first study the integrable $S={{1}\over{2}}$ TFIM and provide an explicit and exact computation of the OTOC starting from the fully polarized state. We show that this does  have a nonzero asymptote everywhere in the ferromagnetic phase, even as the time-ordered autocorrelator of the magnetization vanishes. We uncover a strikingly rich phenomenology in this model summarized in Table \ref{tab:analytical_expressions}.  We  provide an intuitive physical picture for these results, identifying 
the physical processes underpinning the dynamics of domain walls. A central, one is what we term  dynamical relocalization. This is linked to  the absence of chaos, and it explains the  observed signal in the OTOC. {We further find that the state underpinning the OTOC signal resembles a rotated state with low entanglement. }

Intriguingly, in the analogous classical Ising chain, we find that both the OTOC and  autocorrelator of the magnetization exhibit a nonvanishing late-time signal. This can be transparently traced to a failure of the magnetization to thermalize fully. Here, we find that thermalization is promoted by adding disorder to the couplings, thereby removing translational invariance from the Hamiltonian (but not from the initial state of the quench):    
Now the magnetization at long times vanishes classically.

Analogously, the OTOC in the quantum setting is strongly degraded by disorder. Indeed, the main effect of disorder is to reduce  what we refer to as dynamical constraints on the time evolution: These are the integrals of motion which arise due to  
symmetry and integrability of the Hamiltonian, which render it block diagonal. The sense in which the constraints are degraded is that the blocks of the  disordered Hamiltonian have  sizes of $O(N)$, the size of the system, rather than $O(1)$ in the clean case.   

This in turn suggests that the nonvanishing late-time OTOC is predicated on the nonergodicity arising from the integrability of the TFIM. Indeed, a perturbed nonintegrable Ising magnet, initialized with a fully polarized state at weak transverse field, is doubly proximate to integrability: first, on account of the weakness of the perturbation and, second, as a result of hosting a small density of---and hence weakly interacting---domain walls. The apparently nonvanishing OTOC thus is likely only a prethermal signal, visible on the short to intermediate times to which present day numerical tools are limited. 

In the following, we introduce the model, notation, and dynamical observables. We then give a detailed overview of the phenomenology found for the OTOC in contrast with the autocorrelator of the magnetization. Subsequently, we  illustrate how the underlying physical mechanisms---dynamical domain wall relocalization alongside delocalization---arise.  
The role of nonergodicity is underlined by a discussion of the classical transverse field chain, and we conclude with a discussion of  nonintegrable systems.

\begin{figure}
    \centering
    \includegraphics{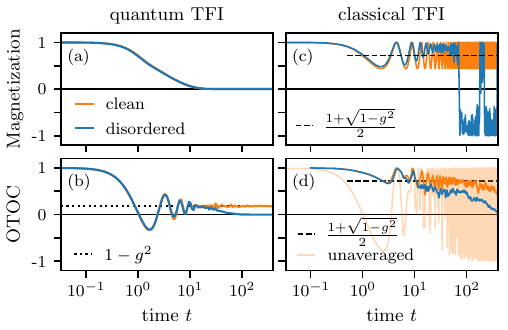}
    \caption{
    Magnetization (top) and OTOC (bottom) for the quantum %\sout
    {(left) and classical (right)} 
    Ising chain given in Eqs.~\eqref{eq:tfi_hamiltonian} and \eqref{eq:classical_model_chain}. We use $N=500$ spins, coupling strength $g=0.9$, open boundary conditions, and $O_\mathrm{m}=O_\mathrm{p}=\sigma^z_{0}$. Static spatial disorder is added to the transverse field (Gaussian with standard deviation $W=0.04$). {(d) shows  results averaged over a time window of $\sim 0.03t$ and over the nine sites closest to the perturbation to reduce fluctuations. For the classical simulations beyond $t \gtrsim 50$, the time evolution of the OTOC appears chaotic.}}
    \label{fig:OP_vs_OTOC_overview}
\end{figure}

\unue{Model and observables. } 
We study the TFIM, a chain of $N$ spins-${1}\over{2}$ (represented by Pauli matrices $\sigma^{x/y/z}_j$) 
\begin{align}
    H_\mathrm{TFI} = -J \sum_{j=-N/2}^{N/2-1} \sigma^z_j \sigma^z_{j+1} + g\sum_{j=-N/2}^{N/2} \sigma^x_j \label{eq:tfi_hamiltonian}
\end{align}
by mapping it to a noninteracting model of fermions corresponding to a $p$-wave superconductor%with spin-triplet pairing
~\cite{pfeuty_order_parameter,Sigrist2005IntroductionTU}. \prbrev{In the following, we set $J=1$.}
%\change{mention non-interacting via JW wherein it looks like p=wave superconductor (since we refer to BCS below)} 
The model has Ising symmetry (generated by $\mathcal{S} = \prod_{i=1}^N \sigma^x_i$), spontaneously broken for $g<1$ via a quantum phase transition at $g=1$. We monitor the magnetization dynamics starting from an initial state $|\Psi\rangle$
\begin{align}
    M_\Psi(t) = \langle \Psi|e^{iHt}\sigma^z_{0} e^{-iHt}|\Psi\rangle \ .
\end{align}
In the thermodynamic limit (TDL), symmetry breaking in the groundstate $|0\rangle$ is diagnosed by nonzero $M_0(t)$.

In addition, for a quench starting with the fully polarized state, a simple OTOC consisting only of tensor product operators $O_\mathrm{m},O_\mathrm{p}$ of Pauli matrices $\sigma^z$
has been proposed to act as a diagnostic of the phase of the ground-state of Eq.~\eqref{eq:tfi_hamiltonian} \cite{heylpollmann,2_X_sun_otoc}
\begin{align}
    C_{O_\mathrm{m},O_\mathrm{p}}(t) = \langle \uparrow^N\!\!|[O_\mathrm{m},O_\mathrm{p}(t)]^2|\!\!\uparrow^N\rangle,
\end{align}
with $O_\mathrm{p}(t)=e^{iHt}O_\mathrm{p} e^{-iHt}$.
Note that $\sigma^\alpha_j(t)$ being Hermitian and unitary implies
\begin{align}
    C_{O_\mathrm{m},O_\mathrm{p}}(t) = 2 \langle \uparrow^N\!\!| O_\mathrm{m} O_\mathrm{p}(t) O_\mathrm{m} O_\mathrm{p}(t) | \!\!\uparrow^N\rangle -2.
\end{align}
The considered OTOC  is the nontrivial part $\mathcal{F}(t)=\langle \uparrow^N\!\!| O_\mathrm{m} O_\mathrm{p}(t) O_\mathrm{m} O_\mathrm{p}(t) | \!\!\uparrow^N\rangle$ of  $C_{O_\mathrm{m},O_\mathrm{p}}(t)$. We call $O_\mathrm{m}$  the measured operator, following the action of a perturbing  $O_\mathrm{p}$.

\unue{Phenomenology in the TFIM. }
Our first central result consists of the time-dependent expectation values of the magnetization and various OTOCs in the TDL, see Figs.~\ref{fig:OP_vs_OTOC_overview} (a) and~\ref{fig:OP_vs_OTOC_overview} (b). This uses the analytical solution of the TFIM and Wick's theorem, as outlined in Refs.~\cite{ lieb61, montrunich_analytical_otoc, calabrese_analyt_decay_mag,silva_annni_and_wick}, with the late-time OTOC values listed in  Table~\ref{tab:analytical_expressions}.

\begin{table}[]
    \centering
    
    \begin{tabular}{l|l|c|c|l}
           $n_p$ & $n_m$ & Reloc. part & Deloc. part & OTOC signal  \\ \hline
          Even & Even & $1$ & $1$ & \multicolumn{1}{c}{$1$} \\
          Even & Odd & $1$ & $\left.\closedsqrt{1-g^2}\right.^{\, n_p}$ & $\left.\closedsqrt{1-g^2}\right.^{\, n_p}$ \\
          Odd & Even & $\left.\closedsqrt{1-g^2}\right.^{\, n_m}$ & $1$ & $\left.\closedsqrt{1-g^2}\right.^{\, n_m}$ \\
          Odd & Odd & $\left.\closedsqrt{1-g^2}\right.^{\, n_m}$ & $\left.\closedsqrt{1-g^2}\right.^{\, n_p}$ & $\left.\closedsqrt{1-g^2}\right.^{\, n_p +n_m}$\\
    \end{tabular}
    \caption{Late-time signals for the different OTOCs in the TDL for the quantum TFIM with $|g|\leq 1$ (for $g>1$, the OTOC signal equals 1 for even/even, and 0 otherwise): $\langle \uparrow^N\!\!\!| O_p(t) O_m O_p(t)|\!\!\uparrow^N\rangle$, with $O_p$ ($O_m$) consisting of a product of $n_p$ ($n_m$) Pauli operators, for concreteness $\sigma^z_{0}\sigma^z_{10}\sigma^z_{20}\cdots$.
    {Note that an odd value of $n_p$ leads to a dependence of the OTOC on $n_m$, and vice versa. A heuristic real-space picture of the corresponding time evolution is given in Fig.~\ref{fig:sketch_relocalisation}.}
    }
    \label{tab:analytical_expressions}
\end{table}

While $M_0(t) \simeq [1-\mathrm{min}(g,1)]^{1/8}$ \cite{pfeuty_order_parameter}, $M_{\uparrow^N}(t)=0$ at late times, see Fig.~\ref{fig:OP_vs_OTOC_overview}, as expected for a state with finite energy density~\cite{heyl_dyn_phase_t,calabrese_analyt_decay_mag,essler_exp_decay,essler_review_domain_walls,silva_annni_and_wick}. By contrast, and 
as noted in Ref.~\cite{heylpollmann}, the OTOCs  mimic an order parameter: They are nonzero for $g<J$ and vanish otherwise. However, note that these are evidently nonthermal: OTOCs which are Ising even, such as those of the bond energy, do not vanish in thermal equilibrium even for $g>J$, {but they do after the quench}. In the remainder of the paper, we  elucidate the physical processes underpinning this nonthermal behavior.

\begin{figure}
                    \centering
                    \includegraphics[scale=0.4]{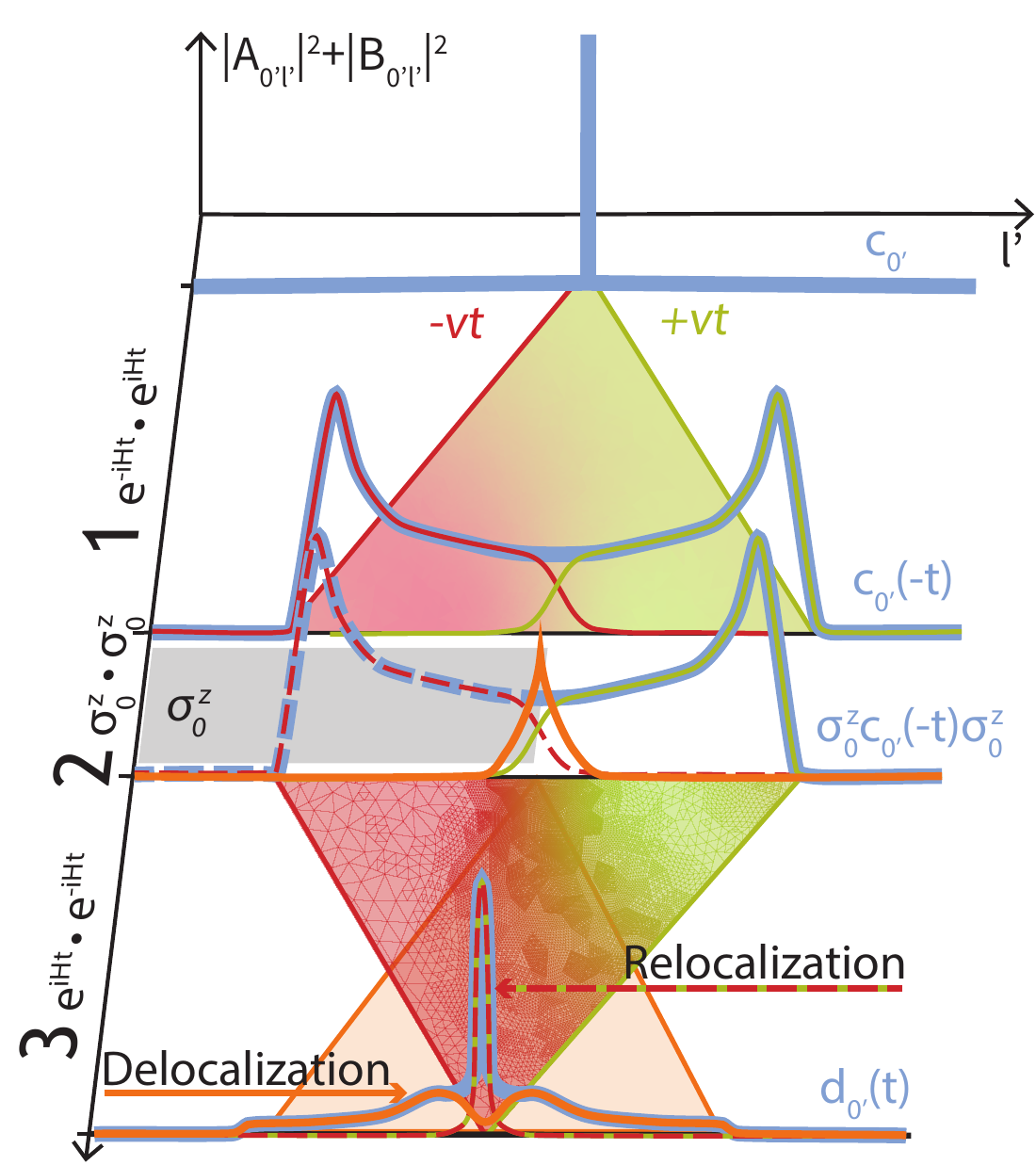}
                    \caption{The schematic creation of $d_{0'}(t)=\sum_{l'} A_{0'l'} c_{l'} + B_{0'l'} c_{l'}^\dagger$, Eq.~\eqref{eq:final_d_opertors}, with relocalization (red, green) and delocalization (orange) in three steps. The blue lines indicate the envelopes of the prefactors $A_{0'l'},B_{0'l'}$ in space when expressing the intermediate stages $c_{0'}$, $c_{0'}(-t)$, $\sigma^z_0 c_{0'}(-t) \sigma^z_{0}$ as sum $A_{0'l'} c_{l'} + B_{0'l'}c_{l'}^\dagger$, respectively. First step: The operator splits into two wavepackets with positive (green) and negative (red) front velocity $v$ creating $c_{0'}(-t)$ . Second step: Applying $\sigma^z_0$ yields a global perturbation by shifting the phase of the left half by $\pi$, indicated by the dashed lines. This decomposes in a phase shift of the left moving wavepacket and a local perturbation shown in orange. Third step: The two wavepackets (green, red) are relocalized, and the orange perturbation delocalizes. Their sum is $d_{0'}(t)$.}
                    \label{fig:sketch_relocalisation}
                \end{figure}

\unue{A domain wall perspective. } 
The above results can be made physically transparent by appealing to the time evolution of domain walls in the Ising chain; for the following heuristic discussion, we refer to the sketch in Fig.~\ref{fig:sketch_relocalisation}. A complementary and detailed formal treatment of relocalization and delocalization is given in the Appendix \ref{app:re_and_deloc}. 

Basically, the exact solution of the Ising chain proceeds via fermionization, in terms of the domain wall annihilation (creation) operators $c_{i'}^{(\dagger)}$ acting on the dual lattice sites at $i'=i-{1\over{2}}$~\cite{montrunich_analytical_otoc,essler_review_domain_walls,santoro_bcs,fisher_tfi_duality}. We find that the dynamics of these operators exhibits two phenomena, which we call domain wall delocalization and relocalization, through which the behavior of the OTOC (and standard observables) can be straightforwardly understood via the following heuristic account. 

The treatment uses the fact that 
the Pauli matrices are unitary operators, so that the solution can be formulated as if studying a time-dependence induced by unitary matrices $\sigma_0^z(t)$. We are thus led to study the time-dependence of  the single fermion operator 
\begin{align}
    d_{j'}(t)= 
    \sigma^z_0(t) c_{j'} \sigma^z_0(t)=\sum_{l'} A_{j'l'} c_{l'} + B_{j'l'} c_{l'}^\dagger,\label{eq:final_d_opertors}
\end{align}
where the second equality follows from $H_\mathrm{TFI}$ being quadratic~\cite{essler_exp_decay,essler_review_domain_walls,montrunich_analytical_otoc} and \prbrev{$\sigma^z_0 c_{i'}^{(\dagger)} \sigma^z_0 = \mathrm{sign}(i') c_{i'}^{(\dagger)}$}.%Eq.~\ref{eq:sigma_z_action}.

\unue{Relocalization and Delocalization. }
Relocalization dominates  the structure of $d_{j'}(t)$ in Eq.~\eqref{eq:final_d_opertors}. It refers to the part of the overlap $\mathrm{Tr}[d_{j'}(t) c^{(\dagger)}_{l'}]$ being time independent and decaying algebraically with $|j'-l'|$, see \prbrev{green and red dashed} line in Fig.~\ref{fig:sketch_relocalisation}. Delocalization is subleading and yields a perturbation of the relocalized part of order ${O}(1/t)$ spread over a region of size $2vt$, with $v$ the maximal group velocity (Fig.~\ref{fig:sketch_relocalisation}, orange line).

The origin of these two phenomena is  sketched in Fig.~\ref{fig:sketch_relocalisation} by dividing the evolution of $d_{0'}(t)$ into three steps: (1)~the evolution of the initial operator $c_{0'}$ with negative time $c_{0'}(-t)$, (2)~the perturbation with $\sigma_0^z$ creating $\sigma^z_0 c_{0'}(-t) \sigma^z_0$ and (3)~a final time evolution $[\sigma^z_0 c_{0'}(-t) \sigma^z_0](t) = d_{0'}(t)$. During the first step, $c_{0'}$ separates in left- and right-moving wave packets. The two wave packets spread ballistically with the maximum group velocity of the domain walls $v$, see red and green lines in Fig.~\ref{fig:sketch_relocalisation}. 
                
Relocalization arises because, in the second step, $\sigma_0^z$ shifts the phase of the left-moving wave packet by $\pi$. The {reverse} time evolution in the third step then cancels with the time evolution of the first, so that both wave packets move back to the origin. Due to the relative phase shift, they no longer add up to $c_{0'}$ but to the relocalized part of $d_{0'}(t)$, \prbrev{green and red dashed} line in Fig.~\ref{fig:sketch_relocalisation}.

The origin of the delocalization also lies in the perturbation with $\sigma^z_0$. For finite times, the wavepackets are not perfectly separated at $l'=0$ and have remaining weight of order ${O}(1/t)$. As a result $\sigma^z_0$ yields a local perturbation of order ${O}(1/t)$, see the orange part in Fig.~\ref{fig:sketch_relocalisation}. In the third step, this locally perturbed part spreads  balistically over a region of size $2vt$. 
Being delocalized, it is not picked up by a fermionic local operator $O_m$ such as $\sigma^z_0\sigma^z_1=1-c_{1'}^\dagger c_{1'}$.
Using purely combinatorial arguments, relocalization and delocalization suffice to explain the structure found in Table~\ref{tab:analytical_expressions}, see Appendix~\ref{app:combinatorial_arguments} and \ref{ssec:app_domain_wall_densities}.

\unue{Classical Ising chain. }
We have found that $\sigma^z(t)|\!\!\uparrow^N\rangle$ represents a lowly entanglement, effectively rotated state, Appendix~\ref{ssec:app_product_state}. It is then natural to consider an entirely classical version of the Ising chain. Here,  we  find the role of   constraints imposed by integrability and symmetry on the observables  to be particularly transparent. The vector of Pauli matrices is replaced by unit vectors $\vec{S}_j$ undergoing precessional dynamics in their net (exchange + applied) field \cite{steinigeweg_quantum_vs_classical,roderich_ballistic_otoc}
\begin{align}\label{eq:classical_model_chain}
    H = - \sum_{j} S_j^z S_{j+1}^z + g \sum_j S^x_j\ .
\end{align}
Due to translational invariance of the initial (fully polarized) state, the equation of motion of $M(t)=S^z_0(t)$---i.e., the zero-wave-vector component of the magnetization--- decouples from the other momentum modes and effectively yields a single-spin problem. The resulting time evolution of $M$ is fully periodic, % \rim{reference figure moved to appendix}
Fig.~\ref{fig:OP_vs_OTOC_overview}(c). Numerically, we find  the amplitude of the periodic motion of the $S^z$ component to be $1-\sqrt{1-g^2}$. 
 
The  classical version of the OTOC forward evolves the fully polarized state, then rotates the spin at site $j$ by $\pi$ around the $z$ axis and finally back-evolves the state. By contrast with $M(t)$, this operation is not spatially uniform, but the state remains polarized with a similar value to $M(t)$, at least for a long initial time, see Fig.~\ref{fig:OP_vs_OTOC_overview}(d).

\unue{Role of disorder. } The stability of even $M(t)$ for the classical case suggests that  constraints on the dynamics---in this case, the decoupling of the uniform $k=0$ mode of the spin density---play a central role in the appearance of nonthermal late-time expectation values. To test this for the classical case, we remove this decoupling by adding to the transverse field Gaussian white disorder with standard deviation $W$. This preserves the Ising symmetry of the system, while the uniformly polarized starting state also remains unchanged. The result is that, already
in the presence of only a small amount of disorder, the classical late-time signal in both $M(t)$  and the OTOC vanishes,  see the right column of Fig.~\ref{fig:OP_vs_OTOC_overview}.

In the quantum system, it is the integrability of the TFIM which yields the (nongeneric) conserved quantities, which follow from the decomposition of the system into 2$\times$2 blocks labeled by the momentum. 
However, adding disorder to the transverse field keeps the single-particle nature of the problem intact while removing the  constraints imposed by translational symmetry and the concomitant momentum conservation. 
Indeed, doing so immediately degrades the signal of the OTOC, Fig.~\ref{fig:OP_vs_OTOC_overview}(b), indicating that the dynamical constraints underpin the late-time signal~\footnote{We note that, while the disorder reduces the dynamical relocalization of domain walls, it instead leads to their Anderson localization, so that for stronger disorder with a short localization length, the OTOC remains visibly nonzero.}. Similarly, as we  argue next, the apparent long-time stability of the quantum OTOC signal in the nonintegrable case  appears to be a finite-time remnant of the integrability of the TFIM and, as such, has the same origin as the relocalization described above.

\unue{Fate of the OTOC beyond integrability. } 
The notion of thermalization states that, in an ergodic system,  observables at a late time after a quench take on the values characteristic of the energy density of the fully polarized state from which the quench started. In the TFIM, this is patently not the case. Concretely, even a local observable such as the exchange energy of neighboring spins $J\, \sigma_j^z\sigma_{j+1}^z$ takes on a nonthermal value: It vanishes for $g>1$, whereas it is in fact nonzero in any thermal state.

What changes  when the constraints imposed by integrability are removed and hence thermalization should occur?
The answer in finite-size {\prbrev{attainable-time} numerics is not much, see Fig.~\ref{fig:compare_annni_dw_density} and Appendix~\ref{ssec:annni_model_study}. The correlators and OTOCs of the ANNNI model described in the next paragraph are very close to those of the integrable TFIM, up to a rescaling of the scale of the effective transverse field. 

This rather suggests that thermalization has simply not taken place in the regime accessible to numerics. 
In fact, we note that the ANNNI model studied in Refs.~\cite{heylpollmann,karrasch_annni,silva_annni_and_wick}, where a second-neighbor interaction is added that breaks integrability $H_\mathrm{ANNNI} = H_\mathrm{TFI} - \Delta \sum_{j} \sigma^z_j\sigma^z_{j+2}$, is close to integrable in two ways: first, due to the smallness of $\Delta$, the integrability-breaking perturbation itself and, second, a four-fermion term of the form 
\begin{align}
    H_\mathrm{int} \sim -4\Delta \sum_j c_{j'}^\dagger c_{j'} c_{j'+1}^\dagger c_{j'+1}. \label{eq:annni_pert}
\end{align}
appears.  This is separately small in the quasiparticle (domain wall) density $c_{j'}^\dagger c_{j'} \sim~g^2/4$. 

The difficulty in confirming definitively that the nonvanishing OTOC is due to a failure to access the actual long-time behavior
lies in the fact that adding a generic perturbation removes integrability and hence makes the problem intractable beyond small finite sizes. However, note that even an Ising-even quantity like the domain wall energy density at long times behaves in essentially the same way as the OTOC in the numerics, see   Fig.~\ref{fig:compare_annni_dw_density}. This would seem to be consistent with the idea that neither quantity has reached its long-time value. As a consequence, either (or perhaps most likely neither) quantity could be used as a ground-state phase diagnostic in the spirit of Ref.~\cite{heylpollmann}: The scattering processes between domain walls are too weak to degrade the signal before finite-size effects mask the behavior of the thermodynamic system. We note, however, that the capacity of various observables to pin down the phase transition point~\cite{2_X_X_titum2019} is not affected by these considerations.

\begin{figure}
    \centering
    \includegraphics{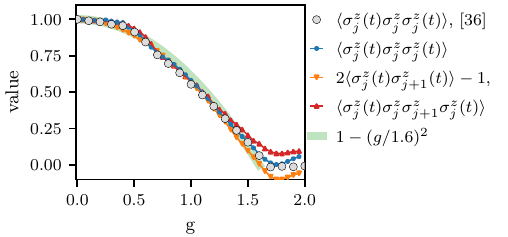}
    \caption{Time-averaged correlators in the nonintegrable ANNNI model at $\Delta=1/2$: OTOC (circles, from Ref.~\cite{heylpollmann} and our own numerics), compared to domain wall density observables (triangles) evaluated in $|\!\!\uparrow^N\rangle$ and $\sigma^z_j(t)|\!\!\uparrow^N\rangle$. The window for time averaging was chosen  after initial transients (by eye), up to the time where boundary effects become visible, \prbrev{see Supplemental Material~\ref{sm:annni_finite}}. The data is most reliable for intermediate values of $g \approx 1$: for small $g$ no clear plateau in the OTOC appears up to $t=120$, while for large $g$, boundary effects set in early.}
    \label{fig:compare_annni_dw_density}
\end{figure}

\unue{Conclusions. } Via a combined numerical and analytical  study of an OTOC in an Ising chain, we have analyzed the interplay of order with many-body dynamics, thermalization and integrability. For the integrable case, where we have identified the mechanism, dynamical domain wall relocalization, underpinning the persistent signal, our treatment is fairly comprehensive; much work remains to fully understand  the behavior of generic many-body systems, here and in higher dimension. 

\begin{acknowledgments}
We are grateful to Fabian Essler, Markus Heyl, Dima Kovrizhin, Adam McRoberts, Benedikt Placke, Frank Pollmann, and Jonathan Nilsson Hallén for engaging discussions. 
This work was in part supported by the Deutsche Forschungsgemeinschaft under Grant SFB 1143 (Project-ID 247310070) and the cluster of excellence ct.qmat (EXC 2147, Project-ID 390858490). 
V.K. was supported in part by the U.S. Department of Energy, Office of Science, Basic Energy Sciences, under Early Career Award No. DE-SC0021111. V.K. also acknowledges support from the Sloan Foundation through a Sloan Research Fellowship and the Packard Foundation through a Packard Fellowship.
\end{acknowledgments}

\newpage

\appendix

\section{Relocalization and delocalization based on Analytical Calculations}\label{app:re_and_deloc}

\subsection{Domain wall description}\label{ssec:app_domain_wall_descr}
In the following, we derive the Gaussian (BCS) formulation of the two states $e^{-iHt}|\!\!\uparrow^N\rangle$ and $\sigma^z_{N/2}(t)|\!\!\uparrow^N\rangle$.

This can be done using a domain wall description. For that, we map the system consisting of $N$ spins on sites $j\in\{1,2,\ldots,N\}$  to a system consisting of $N$ domain walls, i.e., two level systems on the dual lattice with sites $j'\in\{{{1}\over{2}},{{3}\over{2}},\ldots,N-{{1}\over{2}}\}$. A domain wall absent (present) on site $j'$ encodes that the two spins on sites $j'-{{1}\over{2}}$ and $j'+{{1}\over{2}}$ are (anti)aligned. A domain wall absent (present) on site $j'={{1}\over{2}}$ encodes the spin $j=1$ to be the $+1$ ($-1$) eigenstate of $\sigma^z_1$. We describe the absence (presence) of a domain wall on site $j'$ as the $+1$ $(-1)$ eigenstate of $\tau^x_{j'}$. 

In this formulation the reindexed Hamiltonian of Eq.~\eqref{eq:tfi_hamiltonian} reads
\begin{align}
    H_\mathrm{TFI'}= -J \sum_{j'=3/2}^{N-1/2} \tau^x_{j'} + g\sum_{j'=1/2}^{N-3/2} \tau^z_{j'} \tau^z_{j'+1} +  g\tau^z_{N-1/2} \label{eq:dual_h_tfi}
\end{align}
using that $\sigma^z_j\sigma^z_{j+1}\rightarrow \tau^x_{j+1/2}$ and $\sigma^x_j\rightarrow \tau^z_{j-1/2}\tau^z_{j+1/2}$ except for the last spin where $\sigma_N^x\rightarrow \tau^z_{N-1/2}$. Note that the two Hamiltonians Eq.~\eqref{eq:dual_h_tfi} and Eq.~\eqref{eq:tfi_hamiltonian} describe the same matrix.

This Hamiltonian Eqs.~\eqref{eq:dual_h_tfi} cannot be transformed into a local fermionic Hamiltonian due to the last operator $\tau^z_{N-1/2}$. However, if we only intend to measure bulk properties, we can  restrict our calculations to sufficiently small timescales where boundary effects are absent.

A Jordan-Wigner transformation~\cite{santoro_bcs}
\begin{align}
\begin{split}
    \tau^x_{j'} &= (1-2c_{j'}^\dagger c_{j'})\\
    \tau^x_{-N/2-1/2}\cdots \tau^x_{j'-1}\tau^z_{j'} &= (c_{j'}^\dagger + c_{j'}) \end{split} \label{eq:app_jordan_wigner}
\end{align}
yields
\begin{align}
    H_\mathrm{TFI'}=& -J \sum_{j'=3/2}^{N-1/2} (1-2c_{j'}^\dagger c_{j'}) \nonumber \\
    &+ g\sum_{j'=1/2}^{N-3/2} (c_{j'}^\dagger-c_{j'}) (c_{j'+1}^\dagger+c_{j'+1}). \label{eq:hamiltonian_dual}
\end{align}
This Hamiltonian can be diagonalized, as discussed in Ref.~\cite{lieb61} for open boundary conditions and for periodic boundary conditions as discussed in Refs.~\cite{montrunich_analytical_otoc,calabrese_analyt_decay_mag}.

The other operator we are interested in is $\sigma^z_j$. With the duality transformation it transforms to
\begin{align}
    \sigma^z_j = \prod_{j'=-N/2-1/2}^{j-1/2}\tau^x_{j'} \label{eq:sigma_z_domain_walls}
\end{align}
which can be transformed into the fermionic description using Eq.~\eqref{eq:app_jordan_wigner}.

\subsection{Gaussian state formulation}\label{ssec:app_bcs}

We start with a complete set of fermionic operators $c_{j'}$ with pairwise anticommutation relations (ACRs) and their vacuum state with $c_{j'}|0\rangle = 0$. Let $|\Phi\rangle$ be defined as the vacuum of another complete set of fermionic operators with ACRs and $d_{j'} |\Phi\rangle = 0$. Additionally we require the relation
\begin{align}
    d_{j'} = \sum_{l'} A_{j'l'} c_{l'} + B_{j'l'} c_{l'}^\dagger \label{eq:relation_two_vaccuums}.
\end{align}

Then, $|\Phi\rangle$ can be written in Gaussian or BCS type form
\begin{align}\label{eq:fermionic_state_gaussian_form}
    |\Phi\rangle \propto \prod_{j'l'}\left(1+ X_{j'l'} c^\dagger_{j'}c^\dagger_{l'}\right)|0\rangle,
\end{align}
with $2X = - A^{-1} B$ \cite[chapter 5.3]{santoro_bcs}.

\subsection{Domain wall relocalization} \label{ssec:dw_reloc_any_g}
For  ordinary time evolution, $e^{-iH_\mathrm{TFI'}t}|\!\!\uparrow^N\rangle$ is the vacuum of
\begin{align}
    d_{1,j'}(t)= e^{-iH_\mathrm{TFI'}t} c_{j'} e^{iH_\mathrm{TFI'}t}=c_{j'}(-t), \label{eq:definition_first_relation_two_vaccuums}
\end{align}
where we add the index $1$ at $d_{1,j'}$ to distinguish it  from different sets of fermionic operators introduced later.
For this case, the calculation of $A_1,B_1$ in Eq.~\eqref{eq:relation_two_vaccuums} can be performed either for open boundary conditions \cite{lieb61} or for periodic boundary conditions using the analytical solution of $H_\mathrm{TFI'}$ \cite{montrunich_analytical_otoc,calabrese_analyt_decay_mag}. Note that due to inversion symmetry $\mathcal{I}c_{l'}\mathcal{I}^{-1} = i c_{-l'}$ of the Hamiltonian, we have with Eqs.~\eqref{eq:relation_two_vaccuums} and~\eqref{eq:definition_first_relation_two_vaccuums}
\begin{align}
    A_{1,j'l'}=A_{1,(-j')(-l')},  \, B_{1,j'l'}=-B_{1,(-j')(-l')}
\end{align} for $d_{1,j'}(t)=c_{j'}(-t)$. 

For the OTOC, the state $O_\mathrm{p}(t)|\!\!\uparrow^N\rangle$ is the vacuum of
\begin{align}\label{eq:otoc_vacum_op}
    d_{2,j'}(t)= O_\mathrm{p}(t) c_{j'} O_\mathrm{p}(t).
\end{align} 
This can also be written in the form of Eq.~\eqref{eq:relation_two_vaccuums},
as the system is integrable, i.e., an evolved single partice operator remains a single particle operator as can be confirmed by calculating $[c_{j'},H_\mathrm{TFI'}]$ and $[c_{j'},O_\mathrm{p}]$ using \eqref{eq:sigma_z_domain_walls}.
In the following we calculate analytical expressions for $A_{2,j'l'}$ and $B_{2,j'l'}$. We present the calculation in detail below; a rough and simple cartoon of it is provided by the following picture. During the forward evolution part, the domain wall $c_0$ spreads in opposite directions inversion symmetrically, yielding
\begin{align}
    c_0(-t) \approx \alpha(t) (c_{-vt} + c_{vt}) + \beta(t) (-c_{-vt}^\dagger + c_{vt}^\dagger).
\end{align}
Now, the perturbation $O_\mathrm{p}=\sigma^z_0$ yields 
\begin{align}
    \sigma^z_0 c_0(-t) \sigma^z_0 \approx \alpha(t) (-c_{-vt} + c_{vt}) + \beta(t) (c_{-vt}^\dagger + c_{vt}^\dagger).
\end{align}
The back evolution in time preserves the inversion symmetry of the above equation, so that we find to lowest order
\begin{align}
    \sigma^z_0(t) c_0 \sigma^z_0(t) \approx \alpha (c_{1}-c_{-1}) + \beta c_{0}^\dagger
\end{align}
with some prefactors, $\alpha\approx\sqrt{(1-g^2)/2},\beta \approx  g$. Note that here we use the index $0$ to refer to a site in the center of the system.

The analytical calculation now proceeds as follows. We calculate $d_{2,j'}(t)= \sigma^z_{0}(t) c_{j'} \sigma^z_{0}(t)$ using the known solutions for periodic boundary conditions and the formulation for $\sigma^z_0$, Eq.~\eqref{eq:sigma_z_domain_walls}, in the limit $N\rightarrow \infty,t\rightarrow \infty$. We comment in Supplemental Material~\ref{ssec:periodic_vs_open_bc} on calculating Ising odd operators with periodic boundary conditions. 

The Hamiltonian can be diagonalized using a \prbrev{Fourier transformation $c_{j'} = \sum_{k'}e^{ik'j'}/\sqrt{N}$} followed by a Bogoliubov transformation with angle $\tan(\theta_{k'}) = \sin(k')/(g^{-1}+\cos(k'))$, so that $H = \sum \epsilon_{k'}\gamma_{k'}^\dagger \gamma_{k'}$ up to a constant with $\epsilon_{k'}= 2g\sqrt{1+g^{-2}+2g^{-1}\cos k'}$ and $\gamma_k^\dagger = \cos(\theta_k/2)c_{k'}^\dagger -i\sin(\theta_k/2) c_{-k'}$ (having set $J=1$) \cite{montrunich_analytical_otoc,calabrese_analyt_decay_mag}. \prbrev{From this follows that the Hamiltonian acts as a $2\times 2$ subspace spanned by $\gamma_{k'}^\dagger \gamma_{k'}$ being $0$ or $1$ within each momentum subspace.}

Hence, a single propagated domain wall operator is
\begin{align}
    c_{j'}(-t) = \frac{1}{\sqrt{N}}\sum_{k'} &\cos(\theta_{k'}/2)e^{ik'j'} e^{i\epsilon_{k'} t}\gamma_{k'}  \nonumber \\
    &- i\sin(\theta_{k'}/2)e^{ik'j'}e^{-i\epsilon_{k'} t} \gamma_{-k'}^\dagger. \label{eq:single_propagated_domain_wall}
\end{align}
Now we apply the approximation as discussed in the main text. Due to the sign of the group velocity  $\mathrm{sign}(v_{g}(k')) = \mathrm{sign}(\partial_{k'} \epsilon_{k'})=-\mathrm{sign}(k')$, all modes $k'$ separate in space for $t\rightarrow \infty$ according to $\mathrm{sign}(k')$. Essentially, this means that for $c_{j'}(-t)$, for example, the hole part with positive momentum
\begin{align}
    \frac{1}{\sqrt{N}}\sum_{k'>0} &\cos(\theta_{k'}/2)e^{ik'j'} e^{i\epsilon_{k'} t}\gamma_{k'} \nonumber \\
    %c_{j'}(-t) 
    %&= \sum_{k'>0} C_{j'k'}(t) c_{k'} + D_{j'k'}(t) c_{-k'}^\dagger\\
    &= \sum_{l',k'>0} e^{i\epsilon_{k'} t-ik'l'} \left(C_{j'k'} c_{l'} + D_{j'k'} c_{l'}^\dagger\right)
\end{align}
%for example the part 
%\begin{align}
%    \frac{1}{\sqrt{N}} \sum_{k'>0,l'} C_{j'k'}(t) %e^{-ik'l'}c_{l'}
%\end{align}
is located in the left half of the system and
\begin{align}
    %\sum_{l'>0}  \left|\sum_{k'>0} C_{j'k'}(t) e^{-ik'l'}\right|^2 \rightarrow 0\\
    \sum_{l'>0}  \left|\sum_{k'>0} C_{j'k'} e^{i\epsilon_{k'} t-ik'l'}\right|^2 \rightarrow 0
\end{align}
for $t\rightarrow \infty$.

For the choice of $\sigma^z_0$ discussed above, we have $\sigma^z_0 c_{l'} \sigma^z_0 = \mathrm{sign}(l') c_{l'}$ and, hence, can approximate 
\begin{align}\label{eq:reloc_approximation}
    \sigma^z_0 c_{j'}(-t)& \sigma^z_0  \approx \nonumber\\
     \frac{-1}{\sqrt{N}}\sum_{k'}& \mathrm{sign}(k')\cos(\theta_{k'}/2)e^{ik'j'+i\epsilon_{k'} t} \gamma_{k'}  \nonumber \\
    +& \mathrm{sign}(k') i\sin(\theta_{k'}/2)e^{ik'j'-i\epsilon_{k'} t} \gamma_{-k'}^\dagger
\end{align}
which becomes exact for $t\rightarrow \infty$. The relative minus sign stems from the fact that $\gamma_{k'}$ and $\gamma_{-k'}^\dagger$ have opposite group velocities.

Performing the second time evolution on $\sigma^z_0 c_{j'}(-t) \sigma^z_0$ one obtains $d_{2,j'}=\sigma^z_0(t) c_{j'}\sigma^z_0(t)$ with the relation
\begin{align}
    A_{j'l'} = \frac{-1}{N} \sum_{k'} e^{ik'(j'-l')}\mathrm{sign}(k') \cos(\theta_{k'}) \label{eq:momentum_A_bcs_matrix}\\
    B_{j'l'} = \frac{-i}{N} \sum_{k'} e^{ik'(j'-l')}\mathrm{sign}(k') \sin(\theta_{k'}).  \label{eq:momentum_B_bcs_matrix}
\end{align}

Calculating $X_{j'l'}$ can be achieved in momentum space
\begin{align}\label{eq:result_x_rel}
    2X_{j'l'} &= -\sum_{m'} (A^{-1})_{j'm'}B_{m'l'} \nonumber \\
    &= -i\sum_{k'}e^{ik'(j'-l')}\sin(\theta_{k'})/\cos(\theta_{k'}) .
\end{align}
Integrating via the residue theorem yields
\begin{align}
   &\frac{-i}{2\pi} \int_{-\pi}^\pi \! \mathrm{d}k' \frac{\sin(k')}{(g^{-1}+\cos(k'))}e^{i(j'-l')k'}\nonumber \\
   &=-\mathrm{sign}(j'-l')p^{-|j'-l'|}
\end{align}
using that $p^{-1}=e^{ik'_r}$ with $g^{-1}+\cos{k'_r}=0$, and hence, $p^{-1}=-g^{-1}+\sqrt{g^{-2}-1}$. The integration contour is a rectangle, where the integration paths along $-\pi + i \mathrm{Im}(k')$ and $+\pi + i \mathrm{Im}(k')$ cancel as the integrand is $2\pi$ periodic and the integration direction is opposite along the rectangular integration contour. The contributions at $\mathrm{Im}(k')\rightarrow  \infty$ ($-\infty$) vanish as the integrand is exponentially suppressed for $j'-l'>0$ ($j'-l'<0$).

Finally, this exponential decaying $X$ matrix yields a product state as shown in Appendix~\ref{ssec:app_product_state}.

\subsection{A relocalized $X$-matrix yields a product state}\label{ssec:app_product_state}

First, we use that the state is given as
\begin{align}
    \prod_{i'<j'}(1+2X_{i'j'} c_{i'}^\dagger c_{j'}^\dagger) |0\rangle, X_{i'j'} = p^{-|i'-j'|}/2 \label{eq:product_pairs_product_state}
\end{align}
In this state the product generates all possible combinations of products of $O_{i'j'}=c_{i'}^\dagger c_{j'}^\dagger$. For each of these products of pairs of raising operators, the following holds. 

(A) In each  operator product
\begin{align}
    (c_{i'_1}^\dagger c_{j'_1}^\dagger)(c_{i'_2}^\dagger c_{j'_2}^\dagger)\cdots 
\end{align}
each index only occurs once, as operators containing two identical raising operators vanish. 

(B) All pairs of raising operators $c_{i'_1}^\dagger c_{j'_1}^\dagger,\ldots$ commute with all other pairs. This allows us to resort the product of pairs.

(C) Only those operator products remain where there is no overlap between pairs, i.e., given an operator pair $i'_l,j'_l$ in a product of pairs, all other pairs $i'_m,j'_m$ fulfill either 
\begin{align}
    &i'_l < j'_l < i'_m < j'_m \text{ or } i'_m < j'_m < i'_l < j'_l. 
\end{align}
This holds  because, for each operator product where pairs have overlap, there is another operator product generating the same state when applied to $|0\rangle$ with opposite sign. By exchanging for two pairs $i'_1,i'_2$, we get $c^\dagger_{i'_1}c^\dagger_{i'_2}=-c^\dagger_{i'_2}c^\dagger_{i'_1}$ with the same prefactors $X_{i'_1j'_1}X_{i'_2j'_2}$ and $X_{i'_2j'_1}X_{i'_1j'_2}$ as follows from
\begin{align}
    \Leftrightarrow \quad X_{i'_1j'_1}X_{i'_2j'_2} &\stackrel{!}{=} X_{i'_2j'_1}X_{i'_1j'_2}\\
    \Leftrightarrow \quad p^{-|j'_1-i'_1|}p^{-|j'_2-i'_2|} &\stackrel{!}{=} p^{-|j'_1-i'_2|}p^{-|j'_2-i'_1|}\\
    \Leftrightarrow \quad p^{-j'_1+i'_1 - j'_2+i'_2} &\stackrel{!}{=} p^{-|j'_1-i'_2|-|j'_2-i'_1|}\\
    \Leftrightarrow \quad j'_1-i'_1 + j'_2-i'_2 &\stackrel{!}{=} |j'_1-i'_2|+|j'_2-i'_1| \label{eq:condition_prefactors_same}
\end{align}
Now, two pairs having overlap means that $i'_1<j'_2$ and $i'_2<j'_1$, which implies with Eq.~\eqref{eq:product_pairs_product_state} that $j'_1,j'_2$ are both larger than $i'_1$ and $i'_2$. In this case, Eq.~\eqref{eq:condition_prefactors_same} is fulfilled. Hence, the two operator products have the same prefactor $X_{j'_1i'_1}X_{j'_2i'_2} = X_{j'_1i'_2}X_{j'_2i'_1}$. As a result, if two pairs overlap, there is always another operator product in Eq.~\eqref{eq:product_pairs_product_state} with the two operators $c_{i'_1}^\dagger,c_{i'_2}^\dagger$ exchanged which yields a $-1$ sign and these two operator products cancel.

Hence, all operator products generated in  Eq.~\eqref{eq:product_pairs_product_state} cancel which contain two pairs, $i'_1,j'_1$ and $i'_2,j'_2$, fulfilling
\begin{align}
    i'_1<j'_2 \text{ and } i'_2<j'_1.\label{eq:condition_overlap}
\end{align} 
Therefore, only those operator products remain for which Eq.~\eqref{eq:condition_overlap} is not fulfilled for all pairs in the operator product yielding $i'_1>j'_2$ or $i'_2>j'_1$ for all pairs. With $j'_1>i'_1,j'_2>i'_2$ from Eq.~\eqref{eq:product_pairs_product_state}, it implies 
\begin{align}
    j'_1>i'_1>j'_2>i'_2 \text{ or } j'_2>i'_2>j'_1>i'_1,\label{eq:non_overlapping_pairs}
\end{align}
for all pairs, which means that there is no overlap between pairs.
We can verify that those operator products do not cancel, as they have different prefactors: Eqs.~\eqref{eq:non_overlapping_pairs} and~\eqref{eq:condition_prefactors_same} yield
\begin{align}
    \Rightarrow \quad j'_1-i'_1 + j'_2-i'_2 &\stackrel{!}{=} j'_1-i'_2+i'_1-j'_2 \Rightarrow i'_1 = j'_2\\
    (\Rightarrow \quad j'_1-i'_1 + j'_2-i'_2 &\stackrel{!}{=} -j'_1+i'_2-i'_1+j'_2 \Rightarrow j'_1 = i'_2)
\end{align}
contradicting (A). This means that in the case that there is no overlap between two pairs, the operator products related by flipping two operators $i_1',i_2'$ have prefactors which differ in magnitude, so that the operator products do not cancel. 

A unique way of pairing these operator products to cancel respectively is to choose those two overlapping pairs $(j'_1,i'_1),(j'_2,i'_2)$ out of the operator product of which $i'_1$ and $i'_2$ are closest to $N/2$ and in case there is ambiguity, choose that pair with $i'>N/2$ (note that each index can only occur once). Swapping $i'_1\leftrightarrow i'_2$ leaves this choice invariant. Hence, we found, for each operator product with overlapping pairs, a unique second operator product with the opposite prefactor.

Using these properties, we can verify order by order in $p^{-1}$ that the above state coincides with the product state
\begin{align}
    \prod_i (1+p^{-1} \sigma^x_i) |\uparrow^N\rangle.
\end{align}
For that, note that 
\begin{align}
    4 c_{i'}^\dagger c_{j'}^\dagger = \tau^z_{i'}\tau^z_{j'}(\mathds{1}_{i'}+\tau^x_{i'}) \tau^x_{i'+1} \cdots \tau^x_{j'-1} (\mathds{1}_{j'}+\tau^x_{j'}).
\end{align}
This can be transformed using the duality transformation $\tau^z_{i'}\tau^z_{i'+1}=\sigma^x_{i'+1/2}$ and, hence, 
\begin{align}
    \tau^z_{i'}\tau^z_{j'} = (\tau^z_{i'}\tau^z_{i'+1})(\tau^z_{i'+1}\tau^z_{i'+2})\tau^z_{i'+2}\tau^z_{j'} = \sigma^x_{i'+1/2}\cdots \sigma^x_{j'-1/2}\ .
\end{align} 
Similarly, the product of $\tau^x$'s can be transformed into a pair of $\sigma^z$'s yielding
\begin{align}
    4 c_{i'}^\dagger c_{j'}^\dagger = &\sigma^x_{i'+1/2}\cdots \sigma^x_{j'-1/2} \nonumber\\
    &\times(\sigma^z_{i'+1/2}+\sigma^z_{i'-1/2})(\sigma^z_{j'-1/2}+\sigma^z_{j'+1/2}). \label{eq:rewrite_spin_flip_term}
\end{align}
We can now use the properties (A)-(C) when plugging this formula into Eq.~\eqref{eq:product_pairs_product_state}, especially statement (C), where in the expanded form of the product, there are no terms where the $\sigma^x$ parts of the operators in Eq.~\eqref{eq:rewrite_spin_flip_term} overlap or even touch. Hence, for all nonvanishing products, $(\sigma^z_{i'+1/2}+\sigma^z_{i'-1/2})(\sigma^z_{j'-1/2}+\sigma^z_{j'+1/2})=4$, so that we can use

\begin{align}
    (1+2X_{i'j'} c_{i'}^\dagger c_{j'}^\dagger) |0\rangle =  (1+p^{-|i'-j'|} \sigma^x_{i'+1/2}\cdots \sigma^x_{j'-1/2})|0\rangle.
\end{align}
Hence, we create all possible states where the prefactor of each state is given by the product of consecutive spin flips $p^{-|i'-j'|}$. This is the same for the product state
\begin{align}
    \prod_{i}(1+p^{-1}\sigma^x_i)|0\rangle.
\end{align}

\subsection{Delocalization}

The analytical result for the relocalization effect allows for computing the delocalization as the remaining difference
\begin{align}
    X^\mathrm{d} := X - X^\mathrm{r},
\end{align}
with $X^\mathrm{r}$ given in Eq.~\eqref{eq:result_x_rel} and $X$ being the full state as obtained from solving Eq.~\eqref{eq:otoc_vacum_op} and inserting in Eq.~\eqref{eq:fermionic_state_gaussian_form}. Numerical results are shown in Fig.~\ref{fig:decomposition_x_matrix}. This yields the remaining part of the picture presented in the main text: Delocalization following from the wavefunction being not perfectly separated at the moment of applying $O_\mathrm{p}$ as used for the calculation of $X^\mathrm{r}$ in Eq.~\eqref{eq:reloc_approximation}. This delocalized part then spreads ballistically in time with a constant shape, as shown in the inset. This is what is shown schematically in Fig.~\ref{fig:sketch_relocalisation}.

\begin{figure}
    \centering
    \includegraphics{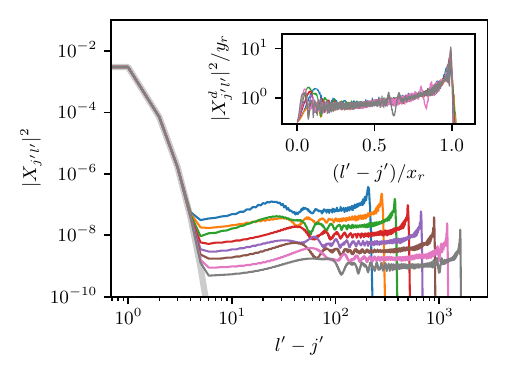}
    \caption{%\rim{drop or move this to SM}
    Central row of $|X_{j'l'}|^2$, with $J = 1, g = 0.3, N=4000$, and times $700<t<5300$. The wide gray line indicates the  exponential decay due to the relocalized domain wall pairs, $X^r$ [Eq.~\eqref{eq:result_x_rel}]. The inset shows its deviation $|X_{j'l'}^d|^2=|X_{j'l'}-X^r_{j'l'}|^2$; these curves collapse upon rescaling the axes with  $x_r=v_\mathrm{wf}t\approx 0.3t$  and $y_r=0.04/t^2$, respectively. The data are averaged over $10$ adjacent $j'$ and two $l'$. %The black line shows \rim{the spatial extent ???}of $c_{j'}(t)$, see also Appendix XXXX.
    }
    \label{fig:decomposition_x_matrix}
\end{figure}

\subsection{Relocalization and delocalization mechanism for multisite operators}\label{app:combinatorial_arguments}

The relocalization mechanism accounts for the results in Table~\ref{tab:analytical_expressions} as follows.
As the left- and right-moving wavefront separate with time, for larger times, every perturbation operator $\sigma^z_0 
\sigma^z_{10} \cdots$ yields a phase shift of the left-moving wavefront by $\pi$ independent of its position. Hence, with an even number of perturbation operators, the relocalization effect does not appear. When relocalization is present, it yields a rotated state on all sites within the light cone, measured by $n_m$ $\sigma^z$ operators reducing the expectation value by $\sqrt{1-g^2}^{n_m}$.

Delocalizing wavefronts are now introduced by each of the perturbation operators separately, as shown for one perturbation operator in Fig.~\ref{fig:sketch_relocalisation}. The delocalized wavefronts reduce the final expectation value of fermionic odd operators by $\sqrt{1-g^2}^{n_p}$, as reported in Table~\ref{tab:analytical_expressions}. Local fermionic even operators do not measure delocalization for large times as they measure local domain wall densities. However, the delocalized part spreads out, see Fig.~\ref{fig:decomposition_x_matrix}, yielding a vanishing domain wall density.

\subsection{Analytical calculation of domain wall densities}\label{ssec:app_domain_wall_densities}
The calculation of 
\begin{align} 
&n_\mathrm{DW} = \nonumber \\
&\lim_{T\rightarrow\infty} \frac{1}{2T}\int_{-T}^T \!\mathrm{d}t \, \frac{1}{N}\sum_{j'} \langle \uparrow^N\!\!|e^{iH_\mathrm{TFI'}t} c_{j'}^\dagger c_{j'} e^{-iH_\mathrm{TFI'}t}|\!\!\uparrow^N\rangle
\end{align}
can be done by using the known transformation into the eigenbasis of $H_\mathrm{TFI'}$ and its inverse. Hence, one obtains in the limit $N\rightarrow \infty$
\begin{align}
    n_\mathrm{DW} &= \frac{1}{4\pi} \int_{-\pi}^\pi \! \mathrm{d}k' \, \sin^2(\theta_{k'}) \\
    &= \frac{1}{4\pi}\int_{-\pi}^\pi \! \mathrm{d}k' \, \frac{\sin^2(k')}{[g^{-1}+\cos(k')]^2 + \sin^2(k')}\\
    &=\begin{cases} \frac{g^2}{4} &\quad g\leq 1 \\\frac{1}{4}  &\quad g>1 \end{cases},
\end{align}
where we used the Weierstrass substitution in the last step to solve the integral.

Using the known relation of $d_{2,j'}$ as given in Eq.~\eqref{eq:momentum_B_bcs_matrix} we can derive
\begin{align}
    &n_\mathrm{DW,OTOC} = \lim_{t\rightarrow \infty} \frac{1}{N} \sum_{j'}\langle \uparrow^N\!\!|\sigma^z_0(t) c_{j'}^\dagger  c_{j'} \sigma^z_0(t)|\!\!\uparrow^N\rangle  \\
   &= \lim_{t\rightarrow \infty} \frac{1}{N} \sum_{j'} \langle \uparrow^N\!\!|\sigma^z_0(t) c_{j'}^\dagger\sigma^z_0(t)\sigma^z_0(t)  c_{j'} \sigma^z_0(t)|\!\!\uparrow^N\rangle\\
   &= \lim_{t\rightarrow \infty} \frac{1}{N} \sum_{j',l'} B_{j'l'}^*B_{j'l'}= \frac{1}{2\pi} \int_{-\pi}^\pi \! \mathrm{d}k' \, \sin^2(\theta_{k'} ) \\
   &= 2n_\mathrm{DW}
\end{align}

This shows that the physics determining the nonanalytic behavior in $n_\mathrm{DW}$ as also discussed in Ref.~\cite{2_X_X_titum2019} and the nonanalytic behavior of the OTOC $n_\mathrm{DW,OTOC}$ are the same: \prbrev{It can be traced back to the band gap closing $\epsilon_{k=\pi}=0$ at $g=1$, yielding the same relation between the OTOC and the band gap closing as found in Ref.~\cite{2_X_X_titum2019} for correlators.}

\subsection{Ising-odd operators and periodic boundary conditions}\label{ssec:periodic_vs_open_bc}
To proceed with the analytical calculation of $A,B$ in Appendix~\ref{ssec:dw_reloc_any_g} we must use periodic boundary conditions, see Refs.~\cite{montrunich_analytical_otoc,calabrese_analyt_decay_mag}. In contrast with the case with open boundary conditions, the special role of the domain wall at site $\frac{1}{2}$ does not play a special role; its presence/absence does not fix the spin at site $1$. Hence, we cannot translate a single $\sigma^z_j$ or, more generally, an Ising-odd $O_\mathrm{p}$ into the fermionic domain wall formulation. Instead, we must use $\sigma^z_j\sigma^z_{j+\lfloor N/2 \rfloor}$, which becomes $\tau^x_{j+1/2}\tau^x_{j+3/2}\cdots \tau^x_{j+\lfloor N/2\rfloor-1/2}$. However, in the TDL, taking first $N\rightarrow \infty$, no domain wall will be influenced within finite times by both, $\sigma^z_j$ and $\sigma^z_{j+\lfloor N/2 \rfloor }$, so that we will only consider one part of the system and use, e.g., $\sigma^z_{j+\lfloor N/2 \rfloor }\rightarrow  \cdots\tau^x_{j+\lfloor N/2 \rfloor -3/2}\tau^x_{j+\lfloor N/2 \rfloor -1/2}$.

\section{OTOC expectation values in the ANNNI model}\label{ssec:annni_model_study}

\prbrev{
\subsection{Quenched OTOCs}\label{sm:annni_finite}
In this section, we provide the numerical results used to generate Fig.~\ref{fig:compare_annni_dw_density}. We simulate the time evolution of the ANNNI model as specified in the main text up to a size of $24$ qubits, see Fig.~\ref{fig:annni_model_sel_time}. By comparing the results for $N=22$ and $24$, we select a window of large times for which boundary effects can still be neglected. We average the OTOC over these times, which yields the data shown in Fig.~\ref{fig:compare_annni_dw_density}.

\begin{figure}[h]
    \centering
    \includegraphics[scale=1]{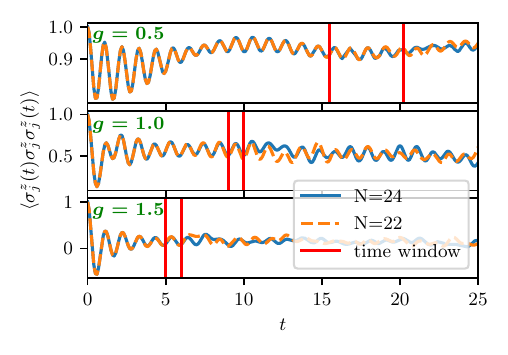}
    \caption{Numerical simulation results for the ANNNI model for $N=22$ (orange) and $N=24$ (blue) for three different exemplary values of $g$, as indicated in green, while $J=1$ and $\Delta=0.5$. The results were obtained using Krylov time evolution. The red lines indicate the manually selected time window over which the OTOC has been averaged, yielding the results shown in Fig.~\ref{fig:compare_annni_dw_density}.}
    \label{fig:annni_model_sel_time}
\end{figure}}

\subsection{Thermal OTOCs}
In this section, we discuss the results for the thermal expectation value for the ANNNI model, taking $\langle \uparrow^N|\sigma^x_0 \sigma^z_0(t)\sigma^x_0 \sigma^z_0(t)|\uparrow^N\rangle$ as a numerically more feasible example. The analytical calculation equivalent to the one outlined in the previous sections yields 
\begin{align}
   \mathrm{Tr}\left[\sigma^x_0 \sigma^z_0(t)\sigma^x_0 \sigma^z_0(t)\right] \stackrel{t\rightarrow\infty}{\rightarrow} \frac{-4}{\pi^2} \begin{cases} 1 & 1 > g\\ \frac{1}{g^2} & 1 \leq g \end{cases}.
\end{align}
for the integrable TFIM for this correlator, yielding a comparable nonanalyticity to $\mathrm{Tr}\left[\sigma^x_0 \sigma^z_0(t)\sigma^x_0 \sigma^z_0(t)\right]$.
The numerical results for the ANNNI model are shown in Fig.~\ref{fig:thermal_annni}. Particulary insightful are the results for $g=1.5,\Delta=0.05$. For this parameter set, the largest bond dimension appears to be sufficient to reliably calculate the signal up to $t\approx 50$ until which the signal has levelled out. This confirms that, also with weak integrability-breaking perturbations, a stable signal at least for intermediate times is observable. The expected long-time decay of the signal due to the quasiparticle interactions cannot be studied reliably due to computational limitations. In general, we observe the required bond dimension to grow rapidly once the OTOC signal decays. For example, for $g=0.5$ or $1.5$ with $\Delta=0.2,\Delta=0.5$ a decay of the signal is clearly visible, but the results are not converged.

\begin{figure}[H]
    %\centering\includegraphics{31_thermal_expectation_value_for_annni_model3.pdf}
    \centering\includegraphics{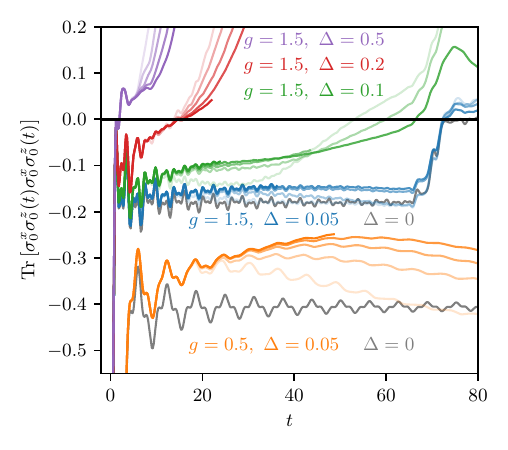}
    \caption{Results for $\mathrm{Tr}\left[\sigma^x_0 \sigma^z_0(t)\sigma^x_0 \sigma^z_0(t)\right]$ for the ANNNI model with $70$ sites and different $\Delta$ and $g$. The simulation was performed using TEBD and second-order trotterization with a step size of $\mathrm{d}t=0.2$. The bond dimension was varied from $64$ (opaque) to $1024$ (solid). For times for which the simulations coincide for different bond dimensions, they also coincide when choosing a smaller $\mathrm{d}t$. The decay of the results for $g=1.5$ at $t\approx 60$ are finite-size effects.}
    \label{fig:thermal_annni}
\end{figure}

\newpage

%\bibliography{refs} % Entries are in the "refs.bib" file

\end{document}